\documentclass[aps,prl,amsmath,amssymb,twocolumn,groupedaddress,
showpacs]{revtex4}

\usepackage{graphicx} 
\usepackage{dcolumn} 
\usepackage{bm}

\catcode`\° = \active 
\def°{\mbox{$^{\circ}$}}

\begin{document}

\title{Measurement of Magnetic-Field Structures in a
  Laser-Wakefield Accelerator}

\author{M.~C.~Kaluza$^{1,2,3}$}
\author{H.-P.~Schlenvoigt$^{1,4}$}
\author{S.~P.~D.~Mangles$^{3}$}
\author{A.~G.~R.~Thomas$^{3,5}$}
\author{A.~E.~Dangor,$^{3}$}
\author{H.~Schwoerer,$^{1,6}$}
\author{W.~B.~Mori,$^{7,8}$}
\author{Z.~Najmudin,$^{3}$}
\author{K.~M.~Krushelnick$^{3,5}$}

\affiliation{$^1$Institut f\"{u}r Optik und
  Quantenelektronik, Friedrich-Schiller-Universit\"{a}t,
  07743 Jena, Germany} 

\affiliation{$^2$Helmholtz-Institut Jena,
  Friedrich-Schiller-Universit\"{a}t, 07743 Jena, Germany}

\affiliation{$^3$Department of Physics, Imperial College,
  London SW7 2AZ, United Kingdom}

\affiliation{$^4$LULI, \'{E}cole Polytechnique, 91128
  Palaiseau cedex, France}

\affiliation{$^5$Center for Ultrafast Optical Science
  (CUOS), University of Michigan, Ann Arbor, Michigan 48109}

\affiliation{$^{6}$Laser Research Institute, Stellenbosch
  University, Stellenbosch 7600, South Africa}

\affiliation{$^{7}$Department of Physics and Astronomy,
  UCLA, Los Angeles, California 90095}

\affiliation{$^{8}$Department of Electrical
  Engineering, UCLA, Los Angeles, California 90095}
  

\begin{abstract}
  Experimental measurements of magnetic fields generated in
  the cavity of a self-injecting laser-wakefield accelerator
  are presented.  Faraday rotation is used to determine the
  existence of multi-megagauss fields, constrained to a
  transverse dimension comparable to the plasma wavelength
  $\sim \lambda_{p}$ and several $\lambda_{p}$
  longitudinally. The fields are generated rapidly and move
  with the driving laser. In our experiment, the appearance
  of the magnetic fields is correlated to the production of
  relativistic electrons, indicating that they are
  inherently tied to the growth and wavebreaking of the
  nonlinear plasma wave. This evolution is confirmed by
  numerical simulations, showing that these measurements
  provide insight into the wakefield evolution with high
  spatial and temporal resolution.
\end{abstract}

\maketitle

The last few years have witnessed tremendous progress in the
field of laser-driven electron acceleration.  Electron beams
with monoenergetic spectra can now be produced with
high-power laser systems \cite{bubble}, potentially
providing a source of brilliant secondary radiation
\cite{sources}. Further improvements have now been made in
terms of energy \cite{gevelectrons} and stability
\cite{stable}.  In these experiments, the high-intensity
laser pulse drives a plasma wave \cite{matlis06}, which
evolves into a `bubble'-like structure with dimensions of
order $\lambda_{p} = 2\pi c /\omega_{p}$ both transversely
and longitudinally \cite{bubble}. The `bubble' compresses
the laser pulse both spatially \cite{selffoc} and temporally
\cite{faure05} provided that it exceeds the power for
relativistic self-focusing $P>P_{cr}\cong 17 (n_{cr}/n_{e})$
GW. This can lead to an increase in laser intensity
eventually leading to wavebreaking and injection of
electrons into the wave's electric field.

This high charge of accelerated electrons ($>$ 100 pC), of
ultrashort bunch duration ($\tau_{b} < \tau_{l} \simeq 50$
fs), thus constitutes an extremely high current ($>$ kA).
Since the current is confined to a dimension much smaller
than the plasma wave ($r \lesssim \mu$m), it will generate a
large associated azimuthal magnetic field, $B_{\varphi} =
\mu_{0}I / 2\pi r$. The field is not neutralized over
distances comparable to the magnetic skin depth
($c/\omega_{p}$) and over associated timescales
($1/\omega_{p}$), which are larger than those of the
electron bunch. Hence these fields will serve to collimate
the beam.

This is not the only magnetic field associated with the
wakefield. In one dimension the coherent motion of electrons
in the plasma wave do not produce a magnetic field, as the
current due to the plasma oscillation $j_{c} = nev$ exactly
cancels the displacement current produced by the resulting
charge separation $j_{d} = \epsilon_{0} \frac{ \partial
  E}{\partial t}$. However in three dimension, electrons are
pushed off-axis and return via the sheath of the `bubble',
so that there is a circulation of current in the wakefield
which generates an azimuthal magnetic field
\cite{magneticwake}. In the `bubble' regime the field can be
calculated by considering the displacement current
associated with the moving electron void $B_\varphi\simeq
\frac{r_{\rm b}}{2} \frac{\partial E_x}{\partial(x-ct)}$
\cite{kostyukov04}, where $r_{\rm b}$ is the blowout radius
of the void and $E_x$ is the longitudinal electric field
\cite{lu06}. Hence the size of the field grows rapidly with
plasma wave amplitude, and thus with increasing laser
intensity. Thus measurements of these fields could provide
important information which is difficult to obtain from
highly non-linear wakefield accelerators \cite{matlis06},
except from their radiation properties \cite{radiation}.

In this Letter, we present the first experimental evidence
for the generation of multi-megagauss azimuthal magnetic
fields in a laser wakefield. The evolution of the magnetic
field indicates the rapid growth and subsequent breaking and
decay of a large amplitude non-linear plasma wave.
Simulations confirm this evolution and show that the field
is a combination of that due to the displacement current of
the plasma wave as well as that associated with the current
of relativistic electrons.

The experiments were performed with the 10-TW Ti:Sapphire
JETI laser at the IOQ, Jena.
\begin{figure}[h]
  \begin{center}
    \includegraphics[width=0.45\textwidth]{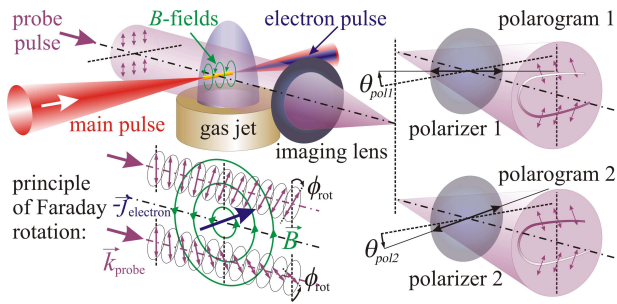}
    \caption{\label{fig:Setup}
      Experimental set-up. An electron beam is generated by
      a self-injecting laser wakefield accelerator. A
      linearly polarized probe pulse traverses the
      interaction region and experiences Faraday rotation.
      The interaction region is imaged onto two CCD-cameras,
      each analysed with a polarizer.}
  \end{center} 
\end{figure}
Pulses of wavelength $\lambda =800$ nm, energy $E_{L} = 800$
mJ, and duration $\tau=85$ fs were focused by an $f/6$
off-axis parabola to intensity $I \simeq
3\times10^{18}\,$Wcm$^{-2}$ onto the rising edge of an
interferometrically characterized helium gas jet from a 1-mm
diameter nozzle (\,Fig.\,\ref{fig:Setup}).  Accelerated
electrons were either detected by a fluorescent screen
shielded by a 30\,$\mu$m Al-foil to block laser light and a
CCD-camera or alternatively by a high-resolution magnetic
spectrometer. Quasi-monoenergetic features peaked around
50\,MeV were observed in the electron spectrum as in
\cite{hidding06}. A synchronized probe pulse was generated
from the transmission of the laser pulse through a 1:100
beam splitter. The probe was telescoped down and guided into
the interaction chamber via a variable delay line
\cite{kaluza08}.  Due to group-velocity dispersion when
propagating through glass, the probe duration was slightly
increased to 100\,fs. The interaction region was backlit by
the probe, which was then split by a non-polarizing beam
splitter and imaged by a high-quality $f/10$ lens with a
magnification of 10 onto a pair of CCD-cameras, equipped
with interference filters.
\begin{figure}[b]
  \begin{center}
    \includegraphics[width=0.4\textwidth]{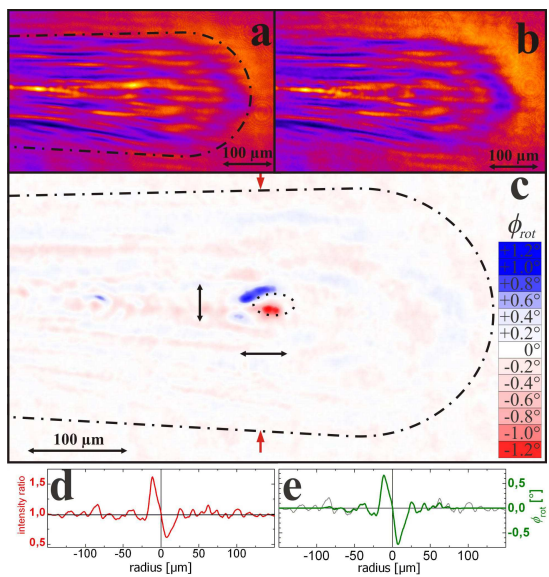}
    \caption{\label{fig:Polarograms}
      Two polarograms in false colour for same laser shot
      with polarizer angles: a) $\theta_{\rm pol1} =
      +5.9^\circ$; b) $\theta_{\rm pol2} = -4.1^\circ$); c)
      distribution of $\phi_{\rm rot}$ deduced from the
      intensity ratio of a) and b). Most pronounced rotation
      occurs in a small region of
      (55$\pm$5)\,$\mu$m$\times$(35$\pm$5)\,$\mu$m indicated
      by black arrows; d) line-out between the red arrows
      shown in c); e) associated $\phi_{\rm rot}$.}
  \end{center} 
\end{figure}
The spatial resolution was 10\,$\mu$m in the transverse and
30\,$\mu$m in the laser direction, determined by the imaging
system and due to the motion blur caused by the orthogonal
motion of the main beam relative to the probe. In addition,
polarizers (extinction ratios of
$(1-\beta_1)=4\times10^{-3}$ and
$(1-\beta_2)=4\times10^{-4}$) were positioned in front of
each camera.  A Nomarskii interferometer \cite{benattar79}
was used to measure the electron density.

The $B_\varphi$-field distribution during the interaction
was measured using Faraday rotation \cite{stamper75}.  When
$B_\varphi$ has components parallel to the propagation of
the probe, the probe polarization is rotated by $\phi_{\rm
  rot}$, (Fig.\,\ref{fig:Setup}). The accumulated rotation
is $\displaystyle \phi_{\rm rot}= \frac{e}{2m_{\rm e}cn_{\rm
    cr}}\int_{\ell} n_{\rm e}(r)\vec{B}_\varphi(r)\cdot{\rm
  d}\vec{s}$ taken along the path $\ell$ of each ray of the
probe through the plasma.  Here, $n_{\rm
  cr}=\varepsilon_0m_{\rm e}\omega_{\rm pr}^2/e^2$ the
critical density for the probe light of frequency
$\omega_{\rm pr}$ and wavenumber $k_{\rm pr}=\omega_{\rm
  pr}/c$. To measure the distribution of $\phi_{\rm rot}$,
the two cameras were carefully aligned to take two images of
the same interaction region.  The two polarizers were
detuned in opposite directions by $\theta_{{\rm pol},1}$ and
$\theta_{{\rm pol},2}$ from the positions crossed with
respect to the initial probe polarization. The intensity
transmitted through polarizer $i$ (where $i=1,2$) can then
be described by Malus' law,
\begin{eqnarray}
  \label{eq:Malus}
  I_{{\rm
      pol,}i}=I_{0}\left[1-\beta_i\sin^2\left[
      90^\circ+\theta_{{\rm pol},i}+\phi_{\rm
        rot}\right]\right],
\end{eqnarray}
where $I_0$ is the initial probe intensity. Taking two
simultaneous polarograms eliminated shot-to-shot
fluctuations of the probe-beam profile including plasma
refraction effects, thus increasing the sensitivity of our
measurements.

Fig.\,\ref{fig:Polarograms}a) and b) show two such
polarograms. The main pulse has entered the plasma from the
left; intensity differences can be seen in the centre of the
images.  This is even more pronounced in the distribution of
$\phi_{\rm rot}$ (Fig.\,\ref{fig:Polarograms}c), deduced
from the ratio of the intensities of the two polarograms
using eq.~(\ref{eq:Malus}). The regions where the strongest
rotation occurs are confined to the center of the image,
being symmetric around the laser axis and having a
longitudinal extent (55$\pm$5)\,$\mu$m and a transverse
diameter of (35$\pm$5)\,$\mu$m. The main pulse was situated
slightly in front of this region, indicated by the dotted
ellipse in Fig.\,\ref{fig:Polarograms}c). The position of
the main pulse was determined by the appearance of ring-like
interference in the images on separate shots where the probe
intensity was reduced to a level comparable to the
side-scattered light from the main pulse. As the
interference can only be generated when the probe and
scattered light overlap in space and time, the center of
these rings marks the instantaneous position of the main
pulse. A lineout of the intensity ratio through this region
and of the associated Faraday-rotation angle are shown in
Fig.\,\ref{fig:Polarograms}d) and e).  The appearance of the
intensity variations was correlated with the generation of
an electron beam.

Two-dimensional (2D) simulations of the interaction were
performed using the particle-in-cell code {\sc osiris}
\cite{fonseca02}. The plasma density profile was set to be
gaussian ($1/e^2$ radius of 550\,$\mu$m) with a peak density
of $4\times10^{19}$cm$^{-3}=0.023 n_{\rm cr}$ as in the
experiment. A laser pulse of gaussian temporal profile with
$\tau = 70$ fs (FWHM), and vacuum focus of 6\,$\mu$m
($1/e^2$ intensity radius) corresponding to
$I=5.4\times10^{18}$\,Wcm$^{-2}$ was set to focus
430\,$\mu$m before the peak density. A slightly higher
initial intensity was chosen since 2D simulations tend to
underestimate effects such as self-focusing which are
important in the experiment. The simulations were performed
in a box size of $800 \times 500\, c/\omega_0$ with
4000$\times$1000 grid cells with 2 macro-particles per cell.
The box moved at the speed of light in the lab frame to
allow higher resolution.

\begin{figure}[h]
  \begin{center}
    \includegraphics[width=0.4\textwidth]{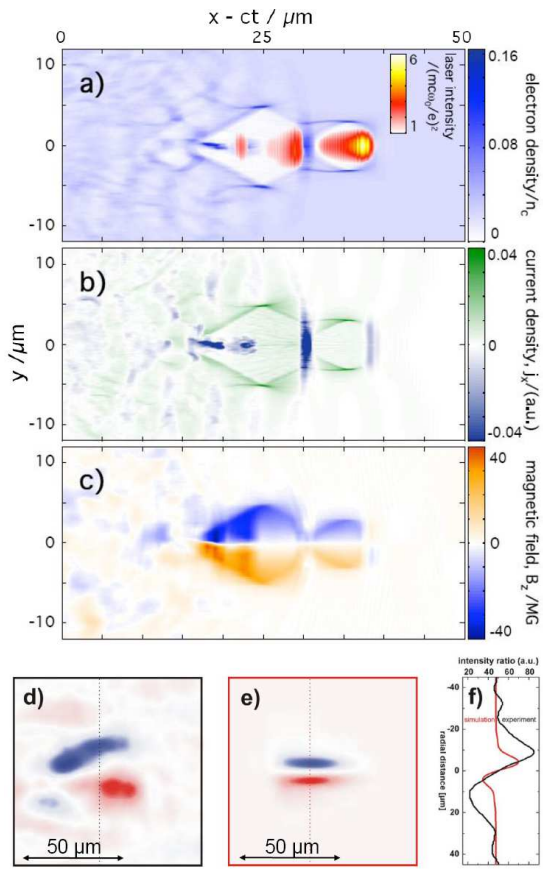}
    \caption{\label{fig:Simulation}
      Simulation results: a) electron density (blue)
      overlaid with laser intensity (red/yellow), b)
      longitudinal current density, c) $B_\varphi$
      d)~92\,$\mu$m$\times92\,\mu$m section of
      Fig.\,\ref{fig:Polarograms}c), e) intensity modulation
      obtained from the numerical simulation including all
      effects affecting the image formation in the
      experiment, f) vertical line-outs along the dotted
      lines in d) and e).}
  \end{center} 
\end{figure}

Fig.\,\ref{fig:Simulation}a) shows the electron density and
the laser intensity from the simulation at the time when
$B_\varphi$ is maximum.
Due to the evolution of the laser, the plasma wake has
evolved from an initially multi-period wave to a
`double-bubble' structure.  Self-modulation of the pulse has
occurred in both the longitudinal and transverse dimensions;
the resulting pulselets contain sufficient energy to drive a
highly non-linear wake structure \cite{hidding06,tsung04}.
Both `bubbles' have a strong $B_\varphi$-field and are of a
similar size, Fig.\,\ref{fig:Simulation}c). The first
`bubble' has a field of magnitude $B_\varphi\sim$15\,MG,
whereas for the second `bubble', $B_\varphi\sim$35\,MG.
However, only the second `bubble' has an on-axis current of
trapped electrons generating a $B_\varphi$-field
(Fig.\,\ref{fig:Simulation}b) indicating that there must be
additional contributions to the field which is present in
both `bubbles'. This is evidently due to the displacement
current.  Both $B_\varphi$ contributions are orientated in
the same way (left handed in direction of laser
propagation).  Even though the peak $B_\varphi$ lies within
the `bubble', where the density is low, in general there is
not complete cavitation and the $B_\varphi$ spreads into the
high-density walls of the `bubble' leading to a measureable
Faraday-rotation signal. Note, however, that the
$B_\varphi$-fields are mainly confined to the `bubble'
structure due to return currents carried by background
electrons flowing around both `bubbles'.

For a quantitative comparison between experiment and
simulation the numerical results shown in
Figs.\,\ref{fig:Simulation}a)-c) were post-processed with a
routine explicitly taking into account experimental effects
influencing the image formation. Based on a ray-tracing
routine accounting for refraction of the probe in the
plasma, we further included the limited spatial resolution
of the imaging system together with the limited temporal
resolution due to the finite probe pulse duration of 100\,fs
and the blurring caused by the perpendicular motion of the
main pulse within the duration of the probe. These effects
serve to increase the apparent size of the feature and also
reducing the signal's amplitude. A comparison between the
$B_\varphi$-field feature's spatial extent and a vertical
lineout of the intensity distribution from the experiment
and the post-processed simulation data are shown in
Figs.\,\ref{fig:Simulation}d)-f). A good agreement is found
both in size and amplitude. For a field of multi-megagauss
strength to be generated by the accelerated electrons,
implies a current of such electrons $>10$ kA. The simulation
shows that such large value of field and current are only
possible because the pulse duration of the electrons is much
shorter than the plasma wave duration (of order $\sim$fs).

\begin{figure}[ht]
  \begin{center}
    \includegraphics[width=0.425\textwidth]{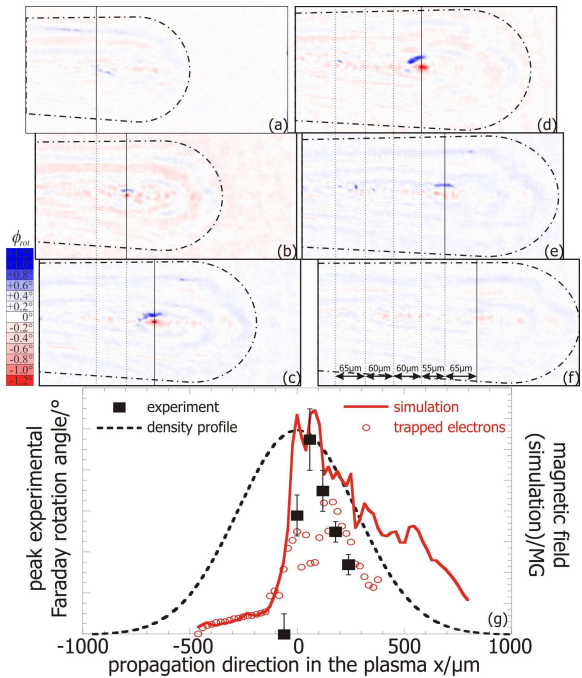}
    \caption{\label{fig:Evolution}
      a)-f) Time evolution of $B_\varphi$. Each image
      (560\,$\mu$m $\times$ 270\,$\mu$m) shows distribution
      of $\phi_{\rm rot}$. g) Peak $\phi_{\rm rot}$
      (squares) as function of position compared to
      corresponding $B_\varphi$ from simulation (red).
      Relative plasma density profile is indicated by dashed
      line. Circles denote $B_\varphi$-contribution from
      trapped electrons. Error determined by difference
      between peak value of $\phi_{\rm rot}$ above and below
      laser axis.}
  \end{center} 
\end{figure}

To separate the contribution to $B_\varphi$ by the wake
displacement current from that from the electron-current,
the individual quasi-particle momenta in the simulation was
recorded.  A particle was considered to be trapped if its
forward momentum $p_x > 7$\,MeV/$c$, corresponding to the
velocity of the plasma `bubble' for $n_{\rm
  e}>9\times10^{18}\,$cm$^{-3}$ and $\lambda_0=800$\,nm
\cite{kostyukov04}. From the current density of all trapped
particles the associated azimuthal magnetic field is
calculated.  It increases rapidly at the same time as the
total $B_\varphi$ confirming that for our experimental
conditions the generation of $B_\varphi$ is strongly
correlated with the trapped electron current.  However, the
accelerated electron alone only accounts for approximately
half of the total $B_\varphi$-field (as shown by the circles
in Fig.\,\ref{fig:Evolution}), confirming the importance of
the $B_\varphi$-field due to the `bubble' itself.

By controlling the delay between main and probe pulse, the
temporal evolution of the magnetic fields was visualized
(Fig.\,\ref{fig:Evolution}a-f). Starting from the first
image where no clear signature can be seen, the feature
appears and reaches a maximum rapidly, but then becomes
weaker again for later times. The distance between the
`bubble'-positions in two subsequent images which were
delayed by 200\,fs stays approximately constant at
$(60\pm5)\,\mu$m leading to a propagation velocity of the
`bubble' of $(3.00\pm0.25)\times10^8$\,ms$^{-1}$, matching
the laser's group velocity in the plasma.

In Fig.\,\ref{fig:Evolution}g), the evolution of the peak
Faraday-rotation angle, which is proportional to the
$B_\varphi$ and which has been measured at different times
in the plasma, is compared to the peak $B_\varphi$ from the
simulation.  After a rapid increase close to the peak of the
density, both quantities show a slower decrease towards the
end of the plasma.
This rapid onset and the subsequent evolution both of the
rotation and the $B_\varphi$-field imply a rapid steepening
and breaking of the plasma wave and subsequent injection of
background electrons into the wake. From these measurements,
it is evident that the wave breaks only once, rather than a
continual injection which is seen in simulations at higher
intensity. The experimental measurements decay faster than
$B_\varphi$ in the simulation. Hence it is apparent that in
the experiment either the confinement of accelerated
electrons and/or the plasma wave amplitude decays much
faster than in the idealized conditions of the simulation.
This is the first visualization of such non-ideal evolution
of a plasma wave, and can help to explain the often large
discrepancies between experiments and simulations.

In conclusion, we have measured the evolution of magnetic
field structures during the interaction of a high-intensity
laser pulse with underdense plasma. This is closely related
to the process of non-linear plasma wave formation and its
breaking into a `bubble'-like structure.  The sudden onset
of the $B_\varphi$-field is a clear signature of the
non-linear steepening of the plasma wave and, for many
experimental conditions, a signature of the subsequent
injection of electrons into the wakefield. By reducing the
plasma density below the threshold for electron
self-injection our technique could also be used to detect
the $B-$fields of the bubble alone. Our diagnostic gives
experimentalists a powerful tool that can easily be
implemented into existing set-ups to provide information of
the acceleration process with high resolution. The results
can directly be compared with numerical simulations which
presently are the basis of our understanding of the
acceleration processes in the plasma, eventually enabling
researchers to identify crucial experimental parameters
required for a stable interaction.  This will help in
optimizing laser-driven electron acceleration, bringing the
prospect of table-top sources of brilliant and ultra-short
XUV and x-rays closer to reality.

This work was supported by DFG (TR18) and BMBF (03ZIK052).
We thank B.~Beleites and F.~Ronneberger for their excellent
laser support.  SPDM thanks the Royal Society and WBM thanks
US DOE and NSF for funding.

\end{document}